\documentclass[draft]{mn2e}
\usepackage{graphicx}
\usepackage{epstopdf}
\usepackage{subfigure}
\usepackage{lscape}
\usepackage{longtable}
\usepackage{wasysym}
\usepackage{psfig}
\usepackage{amssymb}

\usepackage{wasysym}

\newcommand{\counts}{~counts~s$^{-1}$~arcmin$^{-2}$}
\newcommand{\ergs}{~erg~cm$^{-2}$~s$^{-1}$~arcmin$^{-2}$}

\def\gs{\mathrel{\raise0.35ex\hbox{$\scriptstyle >$}\kern-0.6em
\lower0.40ex\hbox{{$\scriptstyle \sim$}}}}
\def\ls{\mathrel{\raise0.35ex\hbox{$\scriptstyle <$}\kern-0.6em
\lower0.40ex\hbox{{$\scriptstyle \sim$}}}}
\def\ls{\mathrel{\hbox{\rlap{\hbox{\lower4pt\hbox{$\sim$}}}\hbox{$<$}}}}
\def\gs{\mathrel{\hbox{\rlap{\hbox{\lower4pt\hbox{$\sim$}}}\hbox{$>$}}}}

\title[Electron Density in Filaments]
      {An Estimate of the Electron Density in Filaments of Galaxies at z $\sim$ 0.1}
\author[A.\ Fraser-McKelvie et al.]
       {Amelia Fraser-McKelvie\thanks{akfra2@student.monash.edu}, Kevin A.\ Pimbblet\thanks{kevin.pimbblet@monash.edu}, and Jasmina S.\ Lazendic\thanks{jasmina.lg@monash.edu}
        \vspace*{1mm}\\
        School of Physics, Monash University, Clayton, VIC 3800, Australia}

\date{\fbox{\sc Draft: \today\ --- Do Not Distribute}}

\pagerange{000--000}

\begin{document}

\maketitle

\begin{abstract}
Most of the baryons in the Universe are thought to be contained within filaments of galaxies, but as yet, no single study has published the observed properties of a 
large sample of {\it known} filaments to determine typical physical characteristics such as temperature and electron density.
This paper presents a comprehensive large-scale search conducted for X-ray emission from a population of 41 bona fide filaments of galaxies 
to determine their X-ray flux and 
electron density. The sample is generated from Pimbblet et al.'s (2004) filament catalogue, which is in turn sourced from the 2 degree Field Galaxy Redshift Survey 
(2dFGRS). Since the filaments are expected to be very faint and of very low density, we used stacked \textsl{ROSAT} All-Sky Survey data. We detect a net surface 
brightness from our sample of filaments of 
$(1.6\pm0.1)\times 10^{-14}$\ergs\ in the 0.9--1.3~keV energy band for 1~keV plasma, which implies an electron density of 
$n_{e} = (4.7\pm0.2)\times 10^{-4}~h_{100}^{1/2}~\rm{cm^{-3}}$. Finally, we examine if a filament's membership to a supercluster leads to an enhanced 
electron density as reported by Kull \& B\"{o}hringer (1999). We suggest it remains unclear if supercluster membership causes such an enhancement.

\end{abstract}


\section{Introduction}

Numerical simulations have shown that a large fraction of the so far unaccounted for baryonic material in the Universe is 
thought to be contained within the Warm-Hot Intergalactic Medium (WHIM) in large-scale, filamentary structures comprised of diffuse gas \cite{Cen,Bregman}. 

Recent wide-field surveys such as the 2dFGRS \cite{Colless} have allowed these 
web-like structures to be probed and the organisation and geometry of filaments connecting cluster pairs to be characterised 
(Pimbblet, Drinkwater \& Hawkrigg 2004). 
Significantly, Arag\'{o}n-Calvo, van de Weygaert \& Jones (2010) suggest that filaments occupy 10 per cent by volume and 39 per cent by mass of the Universe, although this 
figure could be as high as 50 per cent \cite{Bond}. 

While many optical surveys have been successful in detecting and delineating 
filaments in the past (e.g.\ Ebeling, Barret \& Donovan 2004; Pimbblet et al.\ 2004), little work has been 
done on the general X-ray emission from a large sample of these structures, this is not surprising since the individual studies of filaments have shown them to be faint 
structures with low electron density. Filaments are expected to be detected at X-ray wavelengths via thermal bremmstrahlung emission, although some of the first 
detections have been made by observing 0.65~keV O line in absorption or emission. These efforts put into the detection and modeling of the WHIM are summarised by e.g., 
Bregman (2007) and we detail recent X-ray investigations below.

Kull \& B\"{o}hringer (1999) found evidence of filamentary X-ray  emission in the Shapley supercluster. Using data from the \textsl{ROSAT} All-Sky Survey Position 
Sensitive Proportional Counter (PSPC), a diffuse X-ray emission of 17.5 $h_{50}^{-1} \rm{Mpc}$ in length was detected between three clusters and attributed to a filament. The 
Shapley supercluster region is known to have a high concentration of gas and an overdensity of galaxies \cite{Shapley,Raychaudhury,Buote}. Consequently, any filaments 
detected here would be of high density and mass, and their quoted electron density 
would not be indicative of the typical mass distribution of most 
filaments in the Universe, rather a likely upper limit. Taking the turbulent nature of such an overdense region into account, it has been questioned whether the X-ray 
emission detected was indeed evidence of filamentary gas or in fact due to cluster mergers in the region \cite{Scharf}. 

Scharf et al. (2000) was one of the first to detect a filament in both the optical and X-ray bands. Using data from the \textsl{ROSAT} Optical/X-ray survey, 
a low density, diffuse structure of $\sim 12~h_{50}^{-1} \rm{Mpc}$ in length was detected in the CL 1603 field. The observed X-ray emission component of this structure 
was found to have a surface brightness of $\sim1.6 \times 10^{-16}$\ergs\ in the 0.5--2 keV energy band, comparable to estimates for 
intercluster gas.

Tittley \& Henriksen (2001) observed filamentary gas in the region between galaxy clusters Abell 3391 and Abell 3395 using data from the \textsl{ROSAT} and \textsl{ASCA} X-ray satellites. The flux of this gas was found to have the lower limit of 
$1.3 \times 10^{-12}$\ergs\ (0.8--10 keV band) 
associated with a quasi-linear filamentary structure in the line of sight. This result was similar to that derived by Kull \& B\"{o}hringer (1999) for intercluster gas.

More recently, Durret et al. (2003) detected a $4~h^{-1}_{50}$ Mpc long filament using the \textsl{XMM-Newton} X-ray satellite, aligned at the same position 
angle as the major axis of the brightest cluster member. The X-ray 
emission had a temperature of 2 keV, which while cooler than the cluster temperature, is higher than that expected for WHIM. This raises the question of whether this 
emission was from a diffuse filament or ram-stripped gas from an infalling blob.

In a similar vein, Fadda et al. (2008) also detected two filamentary structures between the clusters Abell 1763 and Abell 1770 using the \textsl{Spitzer Space Telescope} in the infra red band. 
A clear detection was made in the 24 $\rm{\mu m}$ band by virtue of a much greater star formation rate within the filament when compared to the background and cluster regions. 
This result supports the concept that galaxies in filaments are indeed falling in toward the clusters they connect \cite{Pimbblet05}. 

Optical and X-ray observations are not the only means of detecting filaments however -- Dietrich et al.(2004) used gravitational lensing to detect the surface mass 
density of a suspected filament between Abell 222 and Abell 223. This detection was backed up by similar detections in the optical and X-ray bands in the same study. 

\begin{figure}
\centerline{\psfig{file=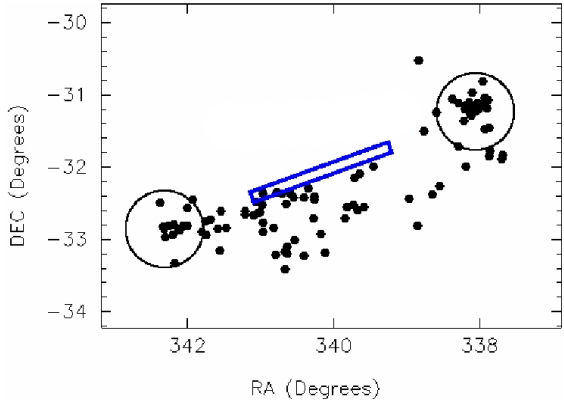,angle=0,width=3.0in}}
  \caption{An example of how a narrow box size could miss most of the mass of a filament. A narrow box is overlaid on the cluster-cluster axis of a `clean', straight filament 
from Porter at al. (2008), the circles denoting the Abell radius of each cluster. It is clear that much of the 
filament is missed by this box, the black dots being galaxies from the 2dFGRS, despite this filament being classified as straight. A larger box shape (as used in the analyses of the current paper) would be more likely to detect the curved filament. }
\label{fig_example}
\end{figure}

While all of the above attempts are single case studies, Briel \& Henry (1995) were the first to conduct a generalised survey of a whole population of 
filaments. Using \textsl{ROSAT} All-Sky Survey (RASS) data, they chose targets between close cluster pairs. While failing to detect evidence of filamentary structure, they 
produced an upper limit on electron density at 7.4 $\times 10^{-5}$ $h_{100}^{1/2} \rm{cm^{-3}} $. 
Briel \& Henry (1995) assumed that all filaments are straight and that every cluster pair would have a filament between them. However, this has been shown not to be 
the case \cite{Pimbblet04,Colberg}. 
For Briel \& Henry's (1995) average filament length of 14.63 $h^{-1}_{100}\rm{Mpc}$, there is only a 86 per cent chance that a 
filament will be connecting the two clusters and a less than 45 per cent chance that this filament will be straight (See Fig. 2 Pimbblet et al. 2004), 
suggesting that perhaps the majority of the filament candidates observed by Briel \& Henry (1995) were either 
non-detections or on the periphery of curved filaments.
It is interesting to note that although all filaments used in this study are classified as `straight' by Pimbblet et al. (2004), there still exists some variation in shape, as an example shown in Figure~\ref{fig_example}.

In this work, we aim to build on previous research in the area by creating a `clean' sample of known straight filaments. Section 2.1 describes the derivation of the sample used, while Section 2.2 gives the method used to obtain the properties of of the filamentary gas. Section 3 presents our results, and Section 4 the discussion and conclusions drawn from this work.

\section{Sample \& Processing}

\subsection{Sample}

Pimbblet et al. (2004) analysed all of the 2dFGRS for cluster pairs and found 805 candidate filaments  of varying geometry. This study draws on the Pimbblet at al. 
(2004) catalogue. Porter et al. (2008) used this list and applied an additional length criterion. Filaments having a length of less than 10 $h_{70}^{-1}$ Mpc were 
rejected because of the difficulty distinguishing these filaments from pockets of gas between merging clusters. From Fig.~2 of Pimbblet et al. (2004) it can be seen 
that the likelihood of a straight filament existing between two clusters separated by a distance larger than 45 $h_{70}^{-1}$ Mpc is low, and hence filaments with a 
length greater than this were rejected from the sample. This left a preliminary sample of 432 filaments between cluster pairs. From these 432 filaments, a `clean' 
sample was created, the conditions of which being: (1) The filaments must be reasonably straight; (2) The filaments must be between two clear clusters and relatively clear of contamination from other filaments and clusters. This reduced the possibility of merger-induced X-ray detections and narrowed the sample to 52 filaments. 

\begin{figure}
\centerline{\psfig{file=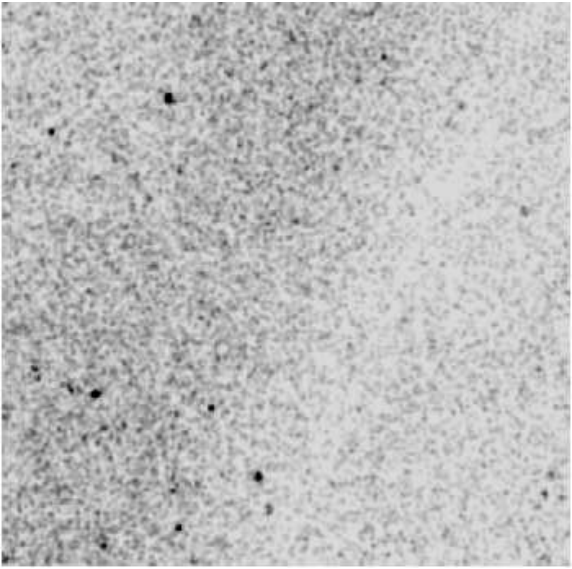,angle=0,width=3.0in}}
  \caption{An example of a smoothed \textsl{ROSAT} frame rejected due to uneven exposure/ high X-ray background variation. Seven such frames containing eleven filaments in total were rejected for this reason.}
\label{fig_variability}
\end{figure}

Of Porter et al.'s (2008) 52 filaments, this study rejected eleven due to uneven exposure time or high variability of the X-ray background in the vicinity, an example of which is shown in Figure~\ref{fig_variability}, 
leaving 41 filaments, listed in Table~\ref{tab_sample}. The galaxy clusters were drawn from the Abell clusters \cite{Abell}, 
the Edinburgh-Durham clusters (EDCC) \cite{Lumsden} and APM clusters (APMCC) \cite{Dalton} catalogues. Redshifts vary from  z = 0.0533 to z = 0.1894, with a mean of 0.0982 and a median of 0.1050. This corresponds to a look back time of $\sim$ 0.9 Gyr, meaning that any evolution effects on the filaments should be minimized, hence the sample is self-similar at high and low z.

\subsection{Data Processing} 

The Max Planck Institute for Extraterrestrial Physics (MPE) archives show X-ray data from the \textsl{ROSAT} All-Sky Survey PSPC instrument available for the intercluster regions of interest. The \textsl{ROSAT} All-Sky Survey data are ideal for this kind of data analysis, as the wide field of view encompasses several clusters and the filaments expected between them.
Two analyses are performed in this work using different energy bands, which are summarized in Table~\ref{tab_analysis}. For the first analysis, we used the RASS data in the 0.5--2.0~keV energy band (channels 52--201), 
where the contribution from the foreground X-ray emission and the detector background are reduced in respect to the full band 0.1--2.3~keV. For the second analysis, we used raw \textsl{ROSAT} event files from MPE archives, processed and filtered to include only channels 91--131 (0.9--1.3~keV), as used by Briel \& Henry (1995), to further minimise possible background contribution.

For each filament, a source extraction region size of $37.5\times37.5$ arcmin$^2$ is chosen to reflect the expected width of the filaments in the line of sight. At a mean redshift of 0.0982, we have a scale 
of 1.273 $h_{100}$ kpc $\rm{arcsec^{-1}}$, corresponding to a $\sim2.9\times 2.9$~$h_{100}$~Mpc box and a 
filament of 1.45~$h_{100}$~Mpc radius. Porter et al. (2008) observations show an average filament width of about twice that used in this paper, suggesting that just the denser middle section of each filament will be detected. The midpoint between the two clusters is calculated and the source counts summed within the region. The background is 
estimated by taking counts in regions (also $37.5\times37.5$ arcmin$^2$ in size) close to the filament, taking care not to include other known filaments or obvious point 
sources. Exposure time for each sample vary from 100~seconds to 700~seconds, with an average value of 285~seconds.

To facilitate a closer comparison with the previous work of Briel \& Henry (1995), we also perform an additional analysis using a narrower energy band (0.9--1.3~keV) but the same size box regions as mentioned above. We choose not to replicate the analysis of Briel \& Henry (1995) exactly by using a narrow, rectangular box aligned with the cluster-cluster axis, as the average box size scaled for the current source sample is such that it would overlap the cluster infall regions on multiple occasions. Also, such a narrow box may likely miss most of the mass of a filament should it be even slightly curved, as demonstrated by Figure~\ref{fig_example}.  As the majority of these clusters are X-ray dark, the optical coordinates from the NASA 
Extragalactic Database (NED) must be used to locate the clusters. Unfortunately, when the optical coordinates are compared to the coordinates of the X-ray visible clusters, quite often they are found to differ from each other, typically by a few arcminutes.
This is not such a problem since the box size that we use means that the filament would still be included in the sample.

\section{Results}

We have not directly imaged extended emission from the filaments in any of the RASS fields used, and hence we proceed with adding the signal from individual sources. In 
Table~\ref{tab_sample} we list the source and background counts obtained from 0.9--1.3~keV band. Listed are the cluster pairs comprising the sample, supercluster Id (Einasto et al.\ 2007), filament Midpoint Right Ascension \& Declination, mean redshift, filament length (L), along with total source counts and total background counts for that filament. In Table~\ref{tab_analysis} 
we summarize the parameters and results from the two analyses using different energy bands, and we compare them to Briel \& Henry's (1995) analysis. 
As analysis 1 uses a greater number of channels, it is expected that it will produce greater values, but may contain more contribution from high energy particles and 
scattered solar flares. Analysis 2 uses a narrower band and an appropriate source box size, and as such, it is believed that it will produce the most robust results. 

The total number of counts in our source and background regions are 640 and 418, respectively, for the 0.9--1.3~keV band (analysis 2). We therefore have some excess photons in our 
source regions over the background regions. The 1$\sigma$ confidence level is thus 20~counts and our 222 excess photons imply $\sim11\sigma$ detection. Taking into account the number of sources, the source region size and the average exposure value of 285~seconds, we find that the filaments have a net flux of $(5.5\pm0.5)\times10^{-4}$\counts\, which translates into a surface brightness of $(1.6\pm0.1)\times10^{-14}$\ergs\ using the $2.89\times10^{-11}$
~erg\,cm$^{-2}$\,s$^{-1}$ conversion factor for Raymond-Smith plasma model with 1~keV temperature, redshift of 0.14 and metallicity of 0.3, as used by Briel \& Henry's (1995). For the 0.5--2.0~keV band (analysis 1), the total number of counts in source and background regions is 4638 and 3732, respectively. Our signal has, therefore, $15\sigma$ significance in this band. Thus, the signal in the 0.5--2.0~keV band is 4 times larger than in the 0.9--1.3~keV band, which is approximately the ratio of the band passes $(201-52)/(131-91) = 3.7$, as expected for a flat photon spectrum and flat effective area.

 The surface brightness $S$ is related to electron density $n_e$ via relation $S = j~n_{e}^{2} L$,  where $L$ is the path length through the filament. For our filament sample we assume $L = 0.7 h_{100}^{-1}$~Mpc.  To derive the electron density Briel \& Henry (1995) used an emission coefficient $j = 6.37 \times 10^{-25}$~erg~cm$^{-3}$~s$^{-1}$~sr$^{-1}$ integrated over the 0.9--1.3~keV band for the above plasma conditions, which gives a following equation
\begin{equation}\label{1}
  n_{e} = 2.01 \times 10^{-2} h^{1/2}_{100} \rm{cm^{-3}} 
\times \sqrt{\rm{PSPC-cnts~s^{-1} arcmin^{-2}} } 
\end{equation}
This gives the filament electron density of $(4.7\pm0.2)\times10^{-4}$~h$_{100}^{1/2}$~cm$^{-3}$ for 0.9--1.3~keV band data.

As mentioned before, the derived values for surface brightness and electron density were taken for the total sample of filaments. 
 When we restrict the sample to include only members of genuine superclusters (4 out of 41 filaments), the net flux of $(2.7\pm0.7)\times10^{-5}$~\counts\  gives surface brighntess of $(7.7\pm2.0)\times10^{-16}$\ergs\ and electron density of $(1.0\pm0.1)\times10^{-4}$~h$_{100}^{1/2}$~cm$^{-3}$, with a detection level of $\sim2.6\sigma$ in 0.9--1.3~keV band. For the 
filaments that only have one cluster as a member of a supercluster (13 of 41 sources), we find the net flux value of $(1.6\pm0.1)\times10^{-4}$\counts, giving a surface brightness of $(4.7\pm0.4)\times10^{-15}$\ergs\ and electron density of $(2.6\pm0.1)\times10^{-4}$~h$_{100}^{1/2}$~cm$^{-3}$ with the detection level of $\sim5\sigma$ in the 0.9--1.3~keV band.

\section{Discussion \& Conclusions}

The similarity in analysis methods between this work and Briel \& Henry (1995) allows a comparison of results, although we use a different filament sample and different box size. Briel \& Henry (1995) quote a net flux of $(4.5\pm4.5) \times 10^{-6}$\counts\ and an upper limit on electron density of $7.4\times 10^{-5}~h_{100}^{1/2}~\rm{cm^{-3}}$. For the same energy band, we obtain flux of $(5.5\pm0.5) \times 10^{-4}$\counts\ and an electron density of $n_e = (4.7\pm0.2)\times 10^{-4}$ h$_{100}^{1/2}$ cm$^{-3}$. 
Our flux is two orders of magnitude higher than their upper limit whilst the electron density is 6 times larger.
The reason we detect a signal is likely due to the fact that this study focused on areas where filaments are known to exist from optical studies, while we assume Briel \& Henry (1995) would most likely have been sampling a random mixture of sources and background counts in their filament sample for which the optical positions of filaments are not known. The reported (and expected) temperature of X-ray plasma in filaments ranges from 0.2--2~keV (see, e.g., Bertone et al.\ 2008), and therefore it is possible to have varying sensitivity to different components of the plasma depending on what energy band is chosen for the analysis. In that regard,  we have chosen two different energy bands for our analysis which gave consistent results.

While  most of our measurements were taken of filaments between clusters that are members of superclusters, this does not take into account the fact that in only one case 
in our sample (Abell 2814 and Abell 2829) 
were these clusters located within the same supercluster. All other cluster pairs are either located within different superclusters to each other, or else one member is 
within a supercluster and another not. This does not give a wide enough sample to definitively state whether the electron  density is higher for filaments contained within superclusters, as the derived values do not differ too much from the total sample, as shown above.

The region encompassing the filament between clusters APMCC 0094 and EDCC 0457 (row 15 of Table~\ref{tab_sample}) has also been studied in detail by other authors.  
Obayashi et al. (2000) used four 
pointed observations from the \textit{ASCA} Gas Imaging Spectrometer (GIS) towards the supercluster containing this filament, but could not find evidence of extended 
emission. 
The filament falls at the 
edge of the GIS detector where sensitivity drops significantly, so indeed, non-detection is not surprising considering the low density derived here. 
Zappacosta et al. (2005) however, analysed 10 fields towards this supercluster and claims a detection of WHIM. Furthermore, they found a correlation between observations 
of gas from the filaments 
as traced by optical data, and soft X-ray emission with T$<$0.5 keV from WHIM. The densities they derived are 10$^{-6}$ cm$^{-3}$ $<$ n $<$ 10$^{-4}$ cm$^{-3}$, and our 
derived values are consistent with their result. We should also note that Sato et al (2010) claims detection of WHIM towards this 
supercluster  using 4 pointings with Suzaku observations, but these observations do not encompass our filament.

As highlighted in the introduction, various methods are employed to try and detect the filaments in individual clusters. Studies of X-ray emission from central regions of 
clusters 
have also provided evidence for the existence of filaments. Bonamente et al. (2002) used 38 pointed \textsl{ROSAT} observations towards the clusters and have found  soft 
X-ray excess at distances larger than 150--200 kpc from the centers of clusters, which could be explained by the presence of filaments. Similarly, Kaastra 
et al. (2003) used pointed \textit{XMM-Newton} observations of 14 clusters to compare column densities derived from X-ray data with those derived from HI 21cm observations 
to separate 5 clusters that show soft X-ray excess with likely origin from WHIM, i.e. filaments.

Finally, our current understanding of the physical conditions within filaments is still mainly provided by simulations. The low densities (10$^{-6}$--10$^{-4}$ cm$^{-3}$) 
and 
high temperatures (10$^{5}$--10$^{7}$ K) expected from such structures in X-rays are presenting an observational challenge. The gas is likely to be inhomogeneous, 
with a range of temperatures and densities, different metallicities and ionization timescales (see Bertone et al. 2008 for a review). The new generation of more sensitive 
X-ray satellites such as \textsl{Suzaku} and upcoming missions should improve the search for such low-density high-temperature 
sources as these filaments (e.g.\ Sato et al.\ 2010).  Indeed, Kawahara et al.\ (2011) have presented evidence for a galaxy group forming at the
intersection of several filaments of galaxies from {\it Suzaku} observations, and further detections from {\it Suzaku} hold much promise in this regard. What is needed now is 
an accurate determination of filaments' electron densities and plasma temperatures from spectral fitting to provide a solid comparison with the models, which will then provide 
more realistic predictions for the missing baryon problem.

Bregman (2007) notes that the observable fraction of the WHIM we seek to detect (i.e.\ T = $10^{6}-10^{7}$ K) should have an 
overdensity of $10^{1}-10^{3}$, which is below current detection thresholds. Hence, our failure to detect individual filaments of galaxies using current methods
is not in conflict with model predictions (Cen \& Ostriker 2006; Bregman 2007 and references therein). 

In summary, we have detected a soft X-ray emission from the combined sample of 41 filaments with known positions using the Pimbblet et al.\ (2004) filament catalogue. We derived an averaged 
electron density for a whole population of filaments of 
$(4.7\pm0.2)\times10^{-4}$~h$_{100}^{1/2}$~cm$^{-3}$ in the 0.9--1.3~keV band, or 
$(9.6\pm4)\times10^{-4}$~h$_{100}^{1/2}$~cm$^{-3}$ in the 0.5--2.0~keV band.

\section*{Acknowledgements}

We thank the referee, J. Henry, for his thorough and insightful comments that have improved the quality of this work. This research has made use of the 
\textsl{ROSAT} All-Sky Survey data which have
been processed at MPE. This research has also made use of the X-Rays Clusters Database (BAX)
which is operated by the Laboratoire d'Astrophysique de Tarbes-Toulouse (LATT),
under contract with the Centre National d'Etudes Spatiales (CNES).

AFM acknowledges the support of a Monash University Summer Vacation Scholarship. 
KAP thanks Somak Raychaudhury for detailed conversations on Filaments over the past 
few years and for hosting him at Birmingham University in late 2009. JSL is supported by a Margaret Clayton Research Fellowship.

\clearpage

\begin{table*}
\caption{The 41 intercluster filaments from the 2dFGRS survey used in this study. The last two columns list the total number of counts per filament in the source and background regions for the 0.9--1.3~keV band (analysis 2). \hfil}
\begin{tabular}{lclcccccccc}
\noalign{\medskip}
\hline 
Cluster 1 & SCl & Cluster 2 & SCl & \multicolumn{2}{c}{Midpoint}                    & Mean     & L                        & Source & Background  \\ 
          &     &           &     & (R.A.~J2000)   & (Dec J2000) & redshift & $(h_{70}^{-1} \rm{Mpc})$ &  (counts)   & (counts) \\
\hline
Abell 2829 & 9 & Abell 0118 & & $00^{h} 53^{m} 35^{s}$ & -27$^{\circ}$27'48'' & 0.1133 & 20.75 & 34 & 12 \\
Abell 2780 & 9 & APMCC 0039 & & $00^{h} 23^{m} 27^{s}$ & -30$^{\circ}$25'50'' & 0.1031 & 33.62 & 14 & 12   \\ 
EDCC 0465 & & Abell 2780 & 9 & $00^{h} 34^{m} 03^{s}$ & -28$^{\circ}$18'04'' & 0.1046 & 40.13 & 7 & 8   \\
Abell 2814 & 9 & Abell 2829 & 9 & $00^{h} 47^{m} 24^{s}$ & -28$^{\circ}$34'12'' & 0.1097 & 22.08 & 4 & 6   \\
Abell 2829 & 9 & EDCC 0511 & & $00^{h} 54^{m} 50^{s}$ & -29$^{\circ}$46'00'' & 0.1112 & 23.01 & 14 & 14   \\
EDCC 0492 & & Abell 2804 & 9 & $00^{h} 47^{m} 01^{s}$ & -29$^{\circ}$58'33'' & 0.1169 & 44.83 & 14 & 6  \\
Abell 1419 & 265 & Abell 1364 & 100 & $11^{h} 49^{m} 58^{s}$ & -1$^{\circ}$00'11'' & 0.1072 & 27.51 & 14 & 12   \\
Abell 1411 & & Abell 1407 & & $11^{h} 54^{m} 25^{s}$ & -1$^{\circ}$08'12'' & 0.1334 & 21.41 & 10 & 7  \\
Abell 3980 & & EDCC 0268 & & $23^{h} 13^{m} 38^{s}$ & -27$^{\circ}$25'43'' & 0.1894 & 27.04 & 10 & 6   \\
Abell 2601 & & Abell 4009 & 299 & $23^{h} 28^{m} 39^{s}$ & -26$^{\circ}$50'58'' & 0.1091 & 42.68 & 10 & 8  \\
EDCC 0365 & & Abell S1155 & & $23^{h} 52^{m} 22^{s}$ & -30$^{\circ}$48'15'' & 0.0547 & 38.71 & 22 & 10   \\
Abell 2741 & & APMCC 0039 & & $00^{h} 15^{m} 40^{s}$ & -31$^{\circ}$58'16'' & 0.1050 & 12.92 & 16 & 10   \\
Abell 2741 & & APMCC 0051 & & $00^{h} 17^{m} 07^{s}$ & -33$^{\circ}$12'04'' & 0.1060 & 21.07 & 24 & 18   \\
EDCC 0445 & & APMCC 0094 & & $00^{h} 32^{m} 54^{s}$ & -29$^{\circ}$11'46'' & 0.0617 & 17.94 & 15 & 11   \\
APMCC 0094 & & EDCC 0457 & & $00^{h} 36^{m} 51^{s}$ & -28$^{\circ}$27'18'' & 0.0613 & 20.86 & 11 & 3   \\
EDCC 0465 & & Abell 2814 & 9 & $00^{h} 41^{m} 05^{s}$ & -27$^{\circ}$55'08'' & 0.1084 & 12.44 & 13 & 7   \\
Abell 2734 & 10 & EDCC 0445 & & $00^{h} 19^{m} 45^{s}$& -28$^{\circ}$13'26'' & 0.0623 & 17.94 & 28 & 22   \\
EDCC 0457 & & EDCC 0445 & & $00^{h} 32^{m} 07^{s}$ & -26$^{\circ}$50'04'' & 0.0620 & 10.24 & 12 & 7  \\
EDCC 0517 & & EDCC 0511 & & $01^{h} 00^{m} 22^{s}$& -32$^{\circ}$10'29'' & 0.1107 & 19.03 & 12 & 10   \\
Abell 2878 & 9 & EDCC 0511 & & $01^{h} 04^{m} 09^{s}$ & -30$^{\circ}$21'06'' & 0.1090 & 24.14 & 14 & 14   \\
Abell 2915 & 28 & EDCC 0581 & & $01^{h} 32^{m} 07^{s}$ & -27$^{\circ}$19'57'' & 0.0862 & 23.51 & 21 & 15   \\
APMCC 0167 & & Abell S0160 & & $01^{h} 26^{m} 51^{s}$ & -32$^{\circ}$55'22'' & 0.0688 & 7.98 & 14 & 12   \\
Abell 2967 & 232$^\star$ & Abell 2972 & 232$^\star$ & $02^{h} 04^{m} 16^{s}$& -27$^{\circ}$41'09'' & 0.1122 & 12.03 & 13 & 7  \\
Abell 2967 & 232$^\star$ & Abell 2981 & 232$^\star$ & $02^{h} 06^{m} 27^{s}$ & -27$^{\circ}$49'29'' & 0.1103 & 16.12 & 18 & 9  \\
Abell 2981 & 232$^\star$ & Abell 2999 & & $02^{h} 13^{m} 41^{s}$ & -27$^{\circ}$52'46'' & 0.1083 & 15.84 & 9 & 8   \\
EDCC 0119 & & Abell 3837 & 190 & $22^{h} 12^{m} 42^{s}$ & -26$^{\circ}$28'05'' & 0.0885 & 28.84 & 14 & 24   \\
EDCC 0057 & & Abell 3837 & 190 & $22^{h} 02^{m} 49^{s}$ & -28$^{\circ}$23'59'' & 0.0922 & 25.07 & 22 & 16   \\
EDCC 0057 & & APMCC 0721 & & $21^{h} 54^{m} 12^{s}$ & -30$^{\circ}$55'31'' & 0.0963 & 30.55 & 22 & 24   \\
Abell 3878 & & Abell 3892 & & $22^{h} 32^{m} 22^{s}$ & -31$^{\circ}$19'33'' & 0.1179 & 22.31 & 18 & 16   \\
EDCC 0230 & & APMCC 0827 & & $22^{h} 54^{m} 38^{s}$ & -31$^{\circ}$00'49'' & 0.1103 & 15.80 & 10 & 4 \\
APMCC 0827 & & Abell 3892 & & $22^{h} 43^{m} 57^{s}$ & -30$^{\circ}$56'27'' & 0.1142 & 31.64 & 24 & 18  \\
EDCC 0128 & & Abell 3880 & 199 & $22^{h} 23^{m} 18^{s}$ & -29$^{\circ}$27'09'' & 0.0586 & 15.12 & 32 & 10   \\
Abell 3959 & & APMCC 0853 & & $23^{h} 04^{m} 08^{s}$ & -32$^{\circ}$04'17'' & 0.0876 & 17.09 & 14 & 2   \\
EDCC 0457 & & Abell 2794 & & $00^{h} 36^{m} 19^{s}$ & -28$^{\circ}$33'33'' & 0.0613 & 20.86 & 7 & 4   \\
EDCC 0202 & & EDCC 0187 & & $22^{h} 48^{m} 31^{s}$ & -33$^{\circ}$23'25'' & 0.0768 & 8.85 & 10 & 6   \\
EDCC 0187 & & APMCC 0810 & & $22^{h} 45^{m} 02^{s}$ & -34$^{\circ}$37'07'' & 0.0769 & 7.24 & 20 & 10   \\
EDCC 0317 & & Abell 4011 & & $23^{h} 32^{m} 49^{s}$ & -34$^{\circ}$02'46'' & 0.1371 & 8.76 & 10 & 6   \\
Abell 3854 & & Abell 3844 & & $22^{h} 15^{m} 37^{s}$ & -35$^{\circ}$14'03'' & 0.1505 & 28.89 & 10 & 4   \\
Abell S1064 & & EDCC 0153 & & $22^{h} 40^{m} 43^{s}$ & -32$^{\circ}$01'46'' & 0.0571 & 17.95 & 24 & 12   \\
EDCC 0239 & & APMCC 0853 & & $23^{h} 04^{m} 42^{s}$ & -32$^{\circ}$11'30'' & 0.0871 & 17.62 & 12 & 2   \\
EDCC 0248 & & Abell 3959 & & $23^{h} 04^{m} 05^{s}$ & -32$^{\circ}$01'36'' & 0.0876 & 17.09 & 18 & 6  \\
\hline 
\multicolumn{6}{l}{$^\star$Candidate supercluster (Einasto et al.\ 2007).} \\
\end{tabular}
\label{tab_sample}
\end{table*}


\begin{table*}
\caption{Comparison of results. Two separate analyses were undertaken in this paper, with different energy bands used. The total summed background and source counts are unknown for Briel \& Henry (1995).}
\begin{tabular}{lccc}
\hline 
   & Present Paper & Present paper & Briel \& Henry (1995) \\
   &  analysis 1    & analysis 2   & \\
\hline
Number of filaments & 41 & 41 & 40$^{\star}$ \\
\textsl{ROSAT} PSPC channel range & 52-201 & 91-131 & 91-131 \\
Energy band (keV) & 0.5--2.0 &  0.9--1.3 & 0.9--1.3 \\
Average exposure time (s) & 285 & 285 & 660 \\
Average redshift    & 0.0982 & 0.0982 & 0.14 \\
Solid angle (arcmin$^2$) & 37.5 $\times$ 37.5 & 37.5 $\times$ 37.5 & 7 $\times$ 114$^{\star\star}$ \\
Average source level (counts) &  4638 & 640 & ? \\
Average background level (counts) & 3732 & 418 & ? \\
Confidence level ($\sigma$) & 15 & 11 & 2 \\
Electron density, $n_{e}$ ($\times10^{-5}(h^{1/2}_{100}\rm{cm}^{-3}$) & $96\pm3$ & $47\pm2$ & $<7.4$ \\
\hline
\multicolumn{4}{l}{$^{\star}$Candidate filaments.} \\
\multicolumn{4}{l}{$^{\star\star}$Averaged solid angle.} \\
 \end{tabular} 
\label{tab_analysis}
\end{table*}


\begin{figure}
\centerline{\psfig{file=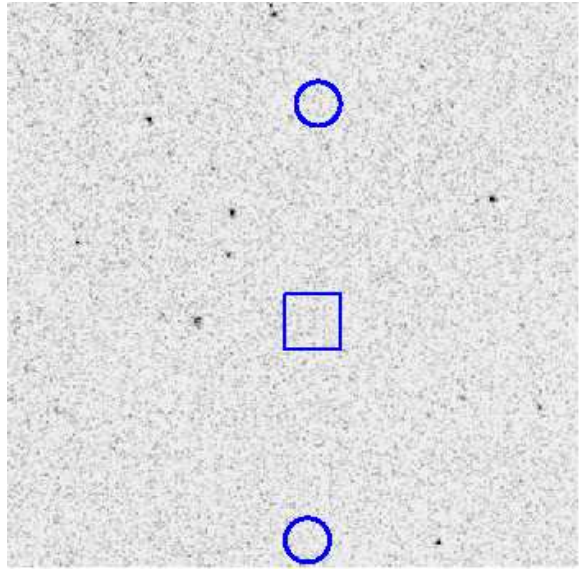,angle=0,width=3.0in}}
  \caption{An example of a smoothed `clean' sample frame (i.e.\ filament is straight, clear of contamination from other filaments and clusters and between two clear clusters), 
with point sources omitted from analysis. The box denotes a filament sample size of 37.5' by 37.5' taken between the 
2 clusters EDCC 0457 (top circle) and Abell 2794 (bottom circle); the field of view is 6.4 $\times$ 6.4 deg$^2$.  
Both of these clusters are X-ray under-luminous and not present in the 
Base de Donn\'{e}es Amas de Galaxies X Database (BAX; Sadat et al.\ 2004), despite being well-known and documented optical galaxy clusters.  Individually, they 
may be under the {\it ROSAT} detection threshold in an analogous manner to our filamentary targets (but also see Gilbank et al.\ 2004).}
\label{fig_clean}
\end{figure}

\end{document}